\documentclass[superscriptaddress,
twocolumn,amsmath,amssymb]{revtex4}
\usepackage[dvips]{graphicx}
\usepackage{epsf}
\usepackage{epsfig}
\usepackage{latexsym}
\usepackage{amssymb}
\usepackage{amsfonts,amsbsy}
\usepackage{amsmath}
\usepackage{dcolumn}
\usepackage{bm}
\usepackage{pifont}
\begin{document}
\title{Traffic of single-headed motor proteins KIF1A: effects of lane changing} 
\author{Debashish Chowdhury}
\email{debch@iitk.ac.in} 
\affiliation{%
Department of Physics, Indian Institute of Technology,
Kanpur 208016, India.
}%
\author{Ashok Garai}
\email{garai@iitk.ac.in} 
\affiliation{%
Department of Physics, Indian Institute of Technology,
Kanpur 208016, India.
}%
\author{Jian-Sheng Wang}%
\email{phywjs@nus.edu.sg}
\affiliation{%
Department of Physics, National University of Singapore, 
Singapore 117542, Singapore 
}%
\date{\today}
\begin{abstract} 
KIF1A kinesins are single-headed motor proteins which move on cylindrical 
nano-tubes called microtubules (MT). A normal MT consists of 13  
protofilaments on which the equispaced motor binding sites form a 
periodic array. The collective movement of the kinesins on a MT is, 
therefore, analogous to vehicular traffic on multi-lane highways 
where each protofilament is the analogue of a single lane. Does 
lane-changing increase or decrease the motor flux per lane? We address 
this fundamental question here by appropriately extending a recent 
model [{\it Phys. Rev. E {\bf 75}, 041905 (2007)}]. By carrying out 
analytical calculations and computer simulations of this extended model, 
we predict that the flux per lane can increase or decrease with the 
increasing rate of lane changing, depending on the concentrations of 
motors and the rate of hydrolysis of ATP, the ``fuel'' molecules. 
Our predictions can be tested, in principle, by carrying out 
{\it in-vitro} experiments with fluorescently labelled KIF1A molecules. 
\end{abstract}
\maketitle

Members of the kinesin superfamily of motor proteins move along 
microtubules (MTs) which are cylindrical nano-tubes 
\cite{schliwa,howard}. A normal MT consists of $13$ protofilaments each 
of which is formed by the head-to-tail sequential lining up of basic 
subunits. Each subunit of a protofilament is a $8$ nm heterodimer of 
$\alpha$-$\beta$ tubulins and provides a specific binding site for a 
single head of a kinesin motor. Often many kinesins move simultaneously 
along a given MT; because of close similarities with vehicular traffic 
\cite{css}, the collective movement of the molecular motors on a MT is 
sometimes referred to as molecular motor traffic 
\cite{polrev,lipo,frey1,santen,popkov}. 

The effects of lane changing on the flow properties of vehicular 
traffic has been investigated extensively using particle-hopping 
models \cite{css} which are, essentially, appropriate extensions of 
the totally asymmetric simple exclusion process (TASEP) 
\cite{sz,derrida,schuetz}. Models of multi-lane TASEP, where the 
particles can occasionally change lane, have also been investigated 
analytically \cite{pronina,hayakawa}. Two-lane generalizations of 
generic models of cytoskeletal molecular motor traffic have also 
been reported \cite{wang07,reichenback}.

Recently a quantitative theoretical model has been developed 
\cite{nosc,greulich} (from now onwards, we shall refer to it as the 
NOSC model) for the traffic of KIF1A proteins, which are single-headed 
kinesins \cite{okada1,okada4,okada5}, by explicitly 
capturing the essential features of the mechano-chemical cycle of each 
individual KIF1A motor, in addition to their steric interactions. 
In this communication we extend the NOSC model by adding to the master 
equation all those terms which correspond to lane changing. Solving 
these equations analytically, we address a fundamental question: does 
lane changing increase or decrease flux per lane? We show that the answer 
to this question depends on the parameter regime of our model. We 
establish the levels of accuracy of our analytical results by comparing 
with the corresponding numerical data obtained from computer simulations 
of the model. We interpret the results physically and suggest experiments 
for testing our theoretical predictions.

\begin{figure}[htb]
\begin{center}
\includegraphics[angle=-90,width=0.9\columnwidth]{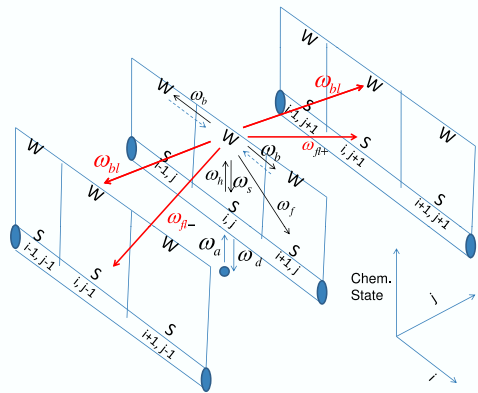}
\end{center}
\caption{(Color online) A schematic representation of our model, 
together will the allowed transitions and the corresponding rate constants.
} 
\label{fig-model}
\end{figure}

The equispaced binding sites for KIF1A on a given protofilament of the 
MT are labelled by the integer index $i$ ($i = 1,...,L$). We use the 
integer index $j$ to label the protofilaments; the position of each 
binding site is denoted completely by the pair $(i,j)$. Because of 
the tubular geometry of the MT, periodic boundary conditions along the 
$j$-direction would be a natural choice. We impose periodic boundary 
conditions also along the $i$-direction, as it not only simplifies 
our analytical calculations but is also adequate to answer the 
fundamental questions which we address in this communication. 

In each mechano-chemical cycle a KIF1A motor hydrolyzes one molecule 
of adenosine triphosphate (ATP) which supplies the mechanical energy 
required for its movement. The experimental results on KIF1A motors 
\cite{okada1,okada4,okada5} indicate that a 
simplified description of its mechano-chemical cycle in terms of a 
2-state model \cite{nosc} would be sufficient to understand their 
traffic on a MT. In the two ``chemical'' states labelled by the 
symbols $S$ and $W$ the motor is, respectively, strongly and weakly 
bound to the MT. 

In the NOSC model, a KIF1A molecule is allowed to attach to (and 
detach from) a site with rates $\omega_{a}$ (and $\omega_{d}$). 
The rate constant $\omega_{b}$ corresponds to the unbiased Brownian 
motion of the motor in the state $W$. The rate constant $\omega_{h}$ 
is associated with the process driven by ATP hydrolysis which causes 
the transition of the motor from the state $S$ to the state $W$. 
The rate constants $\omega_{f}$ and $\omega_{s}$, together, capture 
the Brownian ratchet mechanism \cite{julicher,reimann} of a KIF1A motor. 
Moreover, any movement of the motor under these rules is, finally, 
implemented only if the target site is not already occupied by another 
motor.

The rules of time evolution in the extended NOSC model proposed 
here are identical to those in the NOSC model, except for the 
following additional lane-changing rules (see fig.\ref{fig-model}): \\
a motor weakly-bound (i.e., in state $W$) to the binding site $i$ 
on the protofilament $j$ is allowed to move to the positions $(i,j+1)$ 
and $(i,j-1)$ \\
(i) {\it without} simultaneous change in its chemical state, both the 
corresponding rates being $\omega_{bl}$;\\ 
(ii) {\it with} simultaneous transition to the chemical state $S$,  
the corresponding rate constants being $\omega_{fl+}$ and $\omega_{fl-}$, 
respectively. \\

Let $S_i(j,t)$ and $W_i(j,t)$ denote the probabilities for a motor 
to be in the ``chemical'' states $S$ and $W$, respectively, at site 
$i$ on the protofilament $j$. In the extended NOSC model, under 
mean-field approximation, the master equations for the probabilities 
$S_i(j,t)$ and $W_i(j,t)$  are given by 
\begin{widetext}
\begin{eqnarray}
\frac{dS_i(j,t)}{dt}&=&\omega_a [1-S_i(j,t)-W_i(j,t)] -\omega_h S_i(j,t) -\omega_d S_i(j,t) + \omega_s W_i(j,t) +\omega_f W_{i-1}(j,t)[1-S_i(j,t)-W_i(j,t)]\nonumber\\
&& + \omega_{fl+} [W_{i}(j-1,t)][1-S_{i}(j)-W_{i}(j)] + \omega_{fl-} [W_{i}(j+1,t)][1-S_{i}(j)-W_{i}(j)], 
\label{eq1}
\end{eqnarray}
\begin{eqnarray}
\frac{dW_i(j,t)}{dt}&=& \omega_h S_i (j,t) - \omega_{s} W_{i}(j,t) - \omega_{f} W_{i}(j,t) [1-S_{i+1}(j,t)-W_{i+1}(j,t)]\nonumber\\
&&-\omega_b W_i(j,t)[2-S_{i+1}(j,t)-W_{i+1}(j,t)-S_{i-1}(j,t)-W_{i-1}(j,t)] \nonumber \\
&&+ \omega_b [W_{i-1}(j,t)+W_{i+1}(j,t)][1-S_i(j,t)-W_i(j,t)] \nonumber\\
&&+\omega_{bl} [W_{i}(j-1,t)+W_{i}(j+1,t)][1-S_i(j,t)-W_i(j,t)] \nonumber \\ 
&&- \omega_{bl} W_i(j,t)[2-S_{i}(j+1,t)-W_{i}(j+1,t)-S_{i}(j-1,t)-W_{i}(j-1,t)]\nonumber \\
&&-\omega_{fl+} W_i(j,t)[1-S_{i}(j+1,t)-W_{i}(j+1,t)] - \omega_{fl-} W_i(j,t)[1-S_{i}(j-1,t)-W_{i}(j-1,t)].
\label{eq2}
\end{eqnarray}
\end{widetext}

\begin{table}
\begin{tabular}{|c|c|c|c|c|} \hline
Rate constant&  numerical value/range (s$^{-1}$)\\ \hline
$\omega_{a}$ & 0.1 - 10.0 \\ \hline
$\omega_{d}$ & 0.1 \\ \hline
$\omega_{h}$ & 0 - 250 \\ \hline
$\omega_{s}$ & 145 \\ \hline
$\omega_{f}$ & 55 \\ \hline
\end{tabular}
\caption{\label{tab-parameters}{Rate constants of the NOSC model which 
have been extracted from empirical data on single KIF1A experiments 
\cite{greulich}}.}
\end{table}

\begin{figure}[htb]
\begin{center}
\includegraphics[width=0.9\columnwidth]{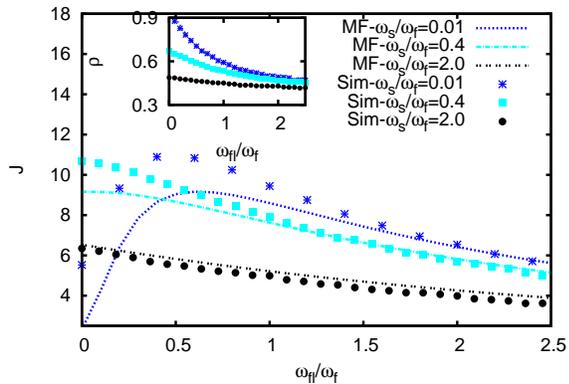}
\end{center}
\caption{(Color online) Flux per lane (and, in the inset, average density of the motors 
on each lane) are plotted against $\omega_{fl}/\omega_{f}$ for 
a few values of $\omega_{s}/\omega_{f}$. Our mean-field predictions 
(labelled MF) are plotted by lines while the discrete data points 
(labelled Sim) have been obtained from our computer simulations of 
the model.  } 
\label{fig-flux1}
\end{figure}

\begin{figure}[htb]
\begin{center}
\includegraphics[width=0.9\columnwidth]{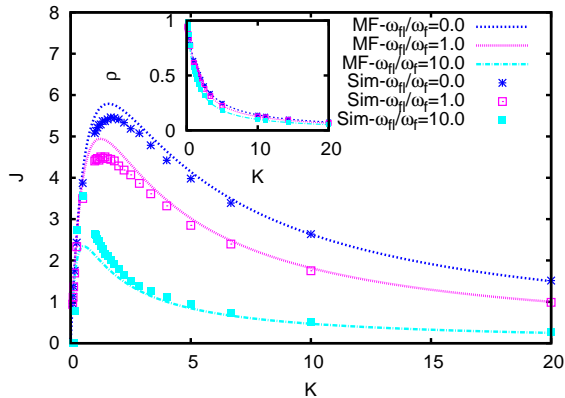}
\end{center}
\caption{(Color online) Same as in Fig.\ref{fig-flux1}, except 
that the data are plotted against $K$ for a few values of 
$\omega_{fl}/\omega_{f}$.} 
\label{fig-flux2}
\end{figure}

\begin{figure}[htb]
\begin{center}
\includegraphics[width=0.9\columnwidth]{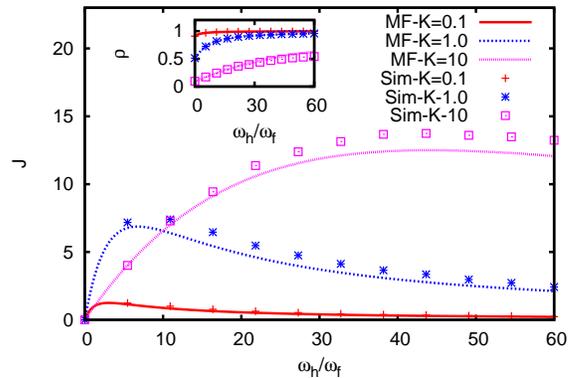}
\end{center}
\caption{(Color online) Same as in Fig.\ref{fig-flux1}, except 
that the data are plotted against $\omega_{h}/\omega_{f}$ for 
a few values of $K$.  } 
\label{fig-flux3}
\end{figure}


In the steady state under {\em periodic} boundary conditions, 
${\tilde S} = S_{i}(j,t)$ and ${\tilde W} = W_{i}(j,t)$, 
independent of $t$ and irrespective of $i$ and $j$; from 
eqs.(\ref{eq1}) and (\ref{eq2}), we get 
\begin{equation}
{\tilde S} = \frac{ -{\tilde{\Omega_h}} - {\tilde{\Omega_s}} - ({\tilde{\Omega_s}} -1)K  +{\sqrt{{\tilde{D}}}}}{2 K(1+K)}
\label{eqS} 
\end{equation}
\begin{equation}
{\tilde W} = \frac{{\tilde{\Omega_h}} +{\tilde{\Omega_s}} + ({\tilde{\Omega_s}} +1)K -{\sqrt{{\tilde{D}}}}}{2 K}, 
\label{eqW}
\end{equation}
where $K=\omega_d/\omega_a$,
${\tilde{\Omega_h}}=\omega_h/{\tilde{\omega_f}}$, ${\tilde{\Omega_s}}=\omega_s/{\tilde{\omega_f}}$, with 
${\tilde{\omega_f}} = \omega_{f}+ \omega_{fl+} + \omega_{fl-}$, and
\begin{equation}
{\tilde{D}}=4{\tilde{\Omega_s}} K(1+K)+
({\tilde{\Omega_h}} +{\tilde{\Omega_s}} + ({\tilde{\Omega_s}}-1)K)^2.
\end{equation}
The average total density of the motors attached to each filament of the 
MT in the steady state is given by
\begin{equation}
\rho = {\tilde S} + {\tilde W} = \frac{{\tilde{\Omega_h}} + {\tilde{\Omega_s}} + ({\tilde{\Omega_s}}+1)K - \sqrt{{\tilde{D}}} + 2}{2(1+K)}.
\label{eq-rho}
\end{equation}

Using the expressions (\ref{eqS}) and (\ref{eqW}) for ${\tilde S}$ 
and ${\tilde W}$, respectively, in the expression 
\begin{equation}
J=\omega_f {\tilde W} (1-{\tilde S}-{\tilde W}) 
\label{eq-flux}
\end{equation}
for the flux of KIF1A motors per lane of the MT highway, we get 
\begin{equation}
J = \frac{\omega_f\biggl[K^2 - \biggl({\tilde{\Omega_h}} + (1+K) {\tilde{\Omega_s}} - \sqrt{{\tilde{D}}}\biggr)^2\biggr]}{4K(1+K)}.
\label{eq-flux2}
\end{equation}

For graphical presentation of our main results, we use the estimates 
of the rate constants, listed in table \ref{tab-parameters}, which 
were extracted earlier \cite{greulich} from empirical data on single 
KIF1A (we use $\omega_{h} = 125$ s$^{-1}$). Since no estimate of 
$\omega_{fl+}$ and $\omega_{fl-}$ are available, we use 
$\omega_{fl+} = \omega_{fl-} = \omega_{fl}$ and vary the single 
parameter $\omega_{fl}/\omega_{f}$ over a wide range to explore the 
consequences of different rates of lane changing.
The flux per lane (obtained from (\ref{eq-flux2})) and the average 
density $\rho$ (given by (\ref{eq-rho})) are plotted against 
$\omega_{fl}/\omega_{f}$ in Fig.~\ref{fig-flux1} for several different 
values of $\omega_{s}/\omega_{f}$ and compared with the corresponding 
simulation data. 

Recall that flux is essentially an average of the product of the density 
and speed of the motors. For sufficiently high $\omega_{s}/\omega_{f}$, 
the density $\rho$ is small even in the absence of lane changing 
($\omega_{fl} = 0$) and, consequently, the motors feel hardly any steric 
hindrance; increasing $\omega_{fl}/\omega_{f}$ in this regime of 
$\omega_{s}/\omega_{f}$ has very little effect on the average speed of 
the motors and it is the decreasing density that is responsible for the 
{\it monotonic} decrease of $J$ with $\omega_{fl}$. 

In sharp contrast, at sufficiently low values of $\omega_{s}/\omega_{f}$, 
$J$ varies {\it non-monotonically} with $\omega_{fl}/\omega_{f}$. In 
this regime of $\omega_{s}/\omega_{f}$, at $\omega_{fl} = 0$, the high 
density of $\rho$ causes steric hindrances which, in turn, leads to 
small $J$. When $\omega_{fl}$ is ``switched on'', $\rho$ decreases 
with increasing $\omega_{fl}$ and $J$ increases up to a maximum because 
of the weakening of the hindrance effects. But, beyond a certain range 
of $\omega_{fl}/\omega_{f}$, the density of motors becomes so low that 
the movement of the motors is practically free of mutual hindrance; 
the decrease of $J$ beyond its maximum is caused by the further reduction 
of density. Larger difference between the predictions of our 
approximate analytical calculations and computer simulation data at 
lower values of $\omega_{s}/\omega_{f}$ arises from the fact that the 
mean-field approximation neglects correlations which increases with 
increasing densitity of the motors.

The above interpretation of trends of variations of $J$ in 
Fig.~\ref{fig-flux1} in terms of the corresponding variation of 
$\rho$ is consistent with the results shown in Figs.~\ref{fig-flux2} 
and \ref{fig-flux3}. But, why does $\rho$ decrease monotonically with 
increasing $\omega_{fl}/\omega_{f}$? Increasing $\omega_{fl}$, keeping 
all the other rate constants unaltered, leads to higher overall rate 
of transitions into strongly-bound states. Since, detachments of the 
motors from the microtubule track take place from the strongly bound 
state (see footnote \cite{note}), the steady-state density is lower 
for higher values of $\omega_{fl}/\omega_{f}$.

An approximate expression for $J$, which is obtained by retaining only 
the terms upto the first order in $\omega_{fl}$ in a Taylor expansion 
of the right hand side of (\ref{eq-flux2}), is given by 
\begin{widetext}
\begin{eqnarray}
J = J_{0} - 4 \biggl(\frac{\omega_{fl}}{\omega_{f}}\biggr) J_{0} 
+ \biggl(\frac{\omega_{fl}}{2\sqrt{D}(1+K)}\biggr)\biggl([(1+K)\Omega_{s}+K]^{2} +[(1+K)\Omega_{s}-\sqrt{D}]^{2}-[\Omega_{h}-\sqrt{D}]^{2} - [\Omega_{h}-K]^{2}\biggr) + O(\omega_{fl}^{2})
\label{eq-flux3}
\end{eqnarray}
\end{widetext}
where $J_{0}$ is the flux corresponding to $\omega_{fl} = 0$ 
(i.e., in the absence of lane changing). The approximate formula 
(\ref{eq-flux3}) still provides a reasonably good estimate of the 
flux even when $\omega_{fl}$ is as large as $\omega_{f}$. 

In this communication we have extended the NOSC model for KIF1A traffic 
on MT \cite{nosc,greulich} by incorporating processes which correspond 
to shifting of the motors from one protofilament to another. These 
processes are analogous to lane changing of vehicles on multi-lane 
highways. On the basis of analytical treatment and computer simulations 
of the extended NOSC model, we have predicted the effects of such 
lane-changing on $J$, the steady-state flux of the KIF1A motors per lane. 
Over a wide region of parameter space, $J$ decreases monotonically with 
increasing value of $\omega_{fl}$, a rate constant for lane-changing. 
However, in some regions of parameter space, $J$ varies  
{\it non-monotonically} with increasing $\omega_{fl}$. We have 
interpreted the results by correlating the observed trends of variation 
of $J$ with the corresponding variation of $\rho$, the average density 
of motors on a lane, and establishing the dependence of $\rho$ on 
$\omega_{fl}$. 

Double-headed conventional kinesin rarely changes lane \cite{ray93}. 
Double-headed dyneins may change lane {\it in-vitro} 
\cite{wang95,peterson06} but, perhaps, not {\it in-vivo} \cite{watanabe07}. 
The bound head of a double-headed motor imposes constraints on the 
stepping of the unbound head. Since such constraints do not exist for 
single-headed kinesins, KIF1A may find it easier to change lane. 
However, KIF1A may dimerize {\it in-vivo} \cite{tomishige}. Therefore, 
{\it in-vitro} experiments with fluorescently labelled KIF1A would be 
able to test our theoretical predictions. In particular, variations of 
$J$ with $K$ and $\omega_{h}$ (see figs.\ref{fig-flux2} and 
\ref{fig-flux3}) can be probed by varying concentration of KIF1A and 
ATP molecules, respectively, in the solution.

\noindent {\bf Acknowledgements}: We thank J. Howard, R. Mallik, 
A. Schadschneider and G.Sch\"utz for useful suggestions. This work 
is supported by the NUS-India Research Initiatives, a Faculty Research 
Grant (NUS), a CSIR research grant (India), physics department of NUS 
and NUS-IITK MoU. 


\begin{thebibliography}{99}
\bibitem{note} The terminology ``strongly-bound '' and ``weakly-bound'' 
states, still used for historical reasons, is misleading; actually, 
the active detachment of KIF1A from the MT track takes place while 
in a transient intermediate state during the transition from the 
``strongly-bound '' state to the ``weakly-bound'' state \cite{okada5}.
\end{thebibliography}
\end{document}